\documentclass[]{elsarticle}
\usepackage{amssymb}
\usepackage{amsthm}

\usepackage{amsmath} 
\usepackage{float}
\graphicspath{{./FIGURES/}}

\begin{document}
\begin{frontmatter}

\title{Epidemics, production and savings. \\Why saving is important?}

\author[label1]{C. F. Gonz\'alez F.}\ead{cf.gonzalez10@uniandes.edu.co}
\author[label1]{J. C. Posada P.}\ead{jc.posada10@uniandes.edu.co}
\author[label2, label3, label4]{J.R. Arteaga B.}\ead{jarteaga@uniandes.edu.co}

\address[label1]{Universidad de los Andes, Facultad de Econom\'ia}
\address[label2]{Universidad de los Andes, Departamento de Matem\'aticas}
\address[label3]{Adjoint at Simon A. Levin Mathematical, Computational \& Modeling Sciences Center, Arizona State University, Tempe, AZ85287, USA}
\address[label4]{Research Group in Mathematical and Computational Biology (BIOMAC) (http://biomac.uniandes.edu.co)}

\begin{abstract}
In this work we  combing models of disease dynamics and economic production, and we show the potential implications of this for demonstrating the importance of savings for buffering an economy during the period of an epidemic. 

Finding an explicit function that relates poverty and the production of a community is an almost impossible task because of the number of variables and parameters that should be taken into account. However, studying the dynamics of an endemic disease in a region that affects its population, and therefore its ability to work, is an honest approach to understanding this function. 
We propose a model, perhaps the simplest, that couples two dynamics, the dynamics of an endemic disease and the dynamics of a closed economy of products and goods that the community produces in the epidemic period.

Some of the results of this study are  expected and known in the literature but some others are not. We highlight three of them: the interdependence that exists between health and the product of that economy.  The majority of the known results only show the dependence in a single direction. Another of the most important goals in this work is to show the existence of a poverty Malthusian trap in economies with low levels of capital investment due to pandemic disease. 
And another  is to show that there is an optimal level of savings that maximizes the \emph{per capita} product of the economy, which is  Machiavellian, but has important implications for market efficiency.
The model tells us why saving is important in an aggregate economy  despite being affected by an epidemic.

\end{abstract}

\begin{keyword}
Infectious diseases; poverty; ecology; economics; complex models
\end{keyword}

\end{frontmatter}

\section{Introduction}
One of the most important and difficult questions for economists is: why do some countries grow while others remain stagnant in poverty? There are many theories that try to explain this phenomenon, such as cultural differences, the role of social institutions or even differences in geographical factors between countries  \cite{REINERT}. However, one possible explanation that has been little studied is that of a poverty trap created by the presence of an infectious disease. According to a WHO study  \cite{WHO1}, infectious diseases are the leading cause of death for the poor. This phenomenon is particularly important in areas such as sub-Saharan Africa, where poor health infrastructure has made it difficult to eradicate dangerous diseases such as malaria  \cite{WHO2}. Since human capital is a key factor in the production of an economy, it is reasonable to assume that the presence of a disease that inhibits the cognitive and motor skills of a population can lead to economic stagnation  \cite{SCHMIDT,DILLON,SACHS,GUY}.

This theory has been studied from several points of view by  Matthew Bonds along with other authors \cite{BONDS1, BONDS2}, in which they use an epidemiological and economic model to model the effects of an infection on the product of an economy and its population levels. In this paper we will start with the epidemiological model used by Bonds and then make a link with a Cobb-Douglas production function, which will model the level of production of the economy based on capital and labor inputs. The objective will be to find the balance of a simplified economy, then find that there is an optimal level of savings that maximizes the \emph{per capita} product of an economy that is subject to the presence of a recurrent  infectious disease.

\section{Epidemiological Model}

The model that will be used to model population and disease evolution is a general SIS (susceptible-infected-susceptible) model  \cite{KERMACK,HETHCOTE1,CCC1,BAILEY,HETHCOTE2}, which describes a disease where individuals can recover and become infected repeatedly over the course of their lives. This model replicates several diseases to which poor communities in areas of the tropics such as malaria are subject. SIS model is one of the simplest epidemic models and is often used to model diseases for which there is no immunity. 

When we speak of disease we are not really talking about a specific disease, but a set of diseases that have similar characteristics. More specifically they are diseases that affect the same population and are recurrent or emerging, are transient or short-duration, are infectious caused by biological agents, such as bacteria, viruses, fungi, etc., are transmitted by vectors or by contagion from one individual to another or by the ingestion of contaminated food or water and in addition not all immunity is acquired. For example, malaria, cancer, dengue, chikungunya, chagas, leishmaniasis, lymphatic filiariasis, sinusitis, pneumonia, meningitis, syphilis, tuberculosis, tetanus,
diarrhea, respiratory diseases and many other neglected diseases.  \cite{WHO1}. Many of these diseases are endemic in central Africa communities.  \cite{JONES,WEBB}

Thus, if an individual acquires for example dengue of serotype 1 (DEN-1), although it acquires immunity for DEN-1 this individual is susceptible to contracting diarrhea or an other disease. That is, an individual can enter/exit to/from compartments of susceptible/infected by  different diseases or the same disease. This is a model assumption which models not a specific disease but a set of diseases. 

\begin{figure}
\centering
\includegraphics[scale=0.6]{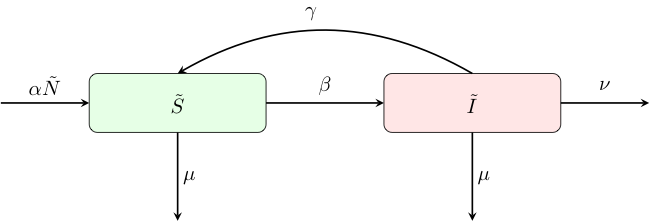}
\caption{Population flow chart of SIS epidemic model.}
\label{CHART_FLOW}
\end{figure}

Fig \ref{CHART_FLOW} shows the flowchart of the SIS model where \(\tilde{S}\) is the susceptible population, \(\tilde{I}\) the population of infected and \(\tilde{N}\) the total population size. The parameter \(\alpha\) will be the birth rate and it is assumed that all individuals enter the susceptible population at birth. The \(\mu\) rate is the natural death rate that applies to both susceptible and infected populations. However, the infected population in addition to dying from natural causes can also die from the disease at a rate \(\nu\). Finally, the \(\beta\) and \(\gamma\) parameters are transmission rate and disease recovery rate respectively.

In this model it is assumed that the total population growth rate is not constant; this means that \(\alpha\) is different from \(\mu + \nu\). It is also assumed that an infected person can relapse into disease indefinitely. Another important assumption of this model and which will be the key to making the coupling with the Cobb-Douglas production function is that the recovery rate is not constant but depends on the health level of each individual which in turn depends on his level of consumption. 
We define this function as follows,

\begin{equation}\label{EQ_1}
\gamma (M) =   \dfrac{\tau M \overline{h} + \kappa \overline{\gamma} }{M + \kappa}
\end{equation}
This function is the result of adding the following two functions,
\begin{equation}\label{EQ_2}
h_{1}(M) = \tau \dfrac{M\overline{h}}{M + \kappa}, \quad 
h_{2}(M) = \dfrac{\kappa \overline{\gamma}}{M + \kappa}
\end{equation}
where \(h(M)\) measures the level of health of the individual which will be a function of income \(M\) \cite{BONDS1}, but we will call the variable \(M\) the level of consumption of the population. In the second part of this manuscript and abusing notation, \(M\) will denote \emph{per-capita} consumption. 
The parameter \(\overline{h}\) is a constant that tells us the maximum possible level of health, while \(\kappa\) is the level of consumption required to obtain the midpoint between the saturation level \(\overline{h}\) of the function \(h(M)\). The value of the \(\overline{\gamma}\) is the value of the typical recovery rate known where there is no consumption, ie for the level of recovery typical of the disease.  The \(\tau\) parameter is an exogenous calibration parameter whose purpose is simply to give the appropriate units so that both functions are rate of change, \(|\tau| = 1\) and we will not write it down.

The recovery function \(\gamma (M)\) is an increasing function between the \(\overline{\gamma}\) and \(\overline{h}\) values. This means that if \(\overline{\gamma}\) is for example \(1/15\), that is to say \(15\) days on average with the illness and \(\overline{h}\) is \(1\) the recovery rate as  a function of consumption level  increases, from a minimum recovery rate of \(1/15\) to a  maximum  of \(1\). That is, the most you can expect is  one day sickness.

Since our SIS model does not have a constant population size \(\tilde{N}\), we will seek the equilibria of the model for susceptible and infected levels when the proportions remain constant. We start from the following equations,

\begin{equation}\label{EQ_3_SYSTEM_1}
\begin{cases}
\dfrac{d\tilde{S}}{dt} = \alpha \tilde{N} + \gamma (M) \tilde{I} - \left(\beta \dfrac{\tilde{I}}{\tilde{N}} + \mu\right) \tilde{S}\\
\dfrac{d\tilde{I}}{dt} = \beta \tilde{S}\dfrac{\tilde{I}}{\tilde{N}} - \left( \mu + \nu + \gamma (M) \right)\tilde{I}
\end{cases}
\end{equation}
which show that the total population size \(\tilde{N}\) do not remain constant w.r.t. time \(t\). Using the change of coordinates \(S = \tilde{S}/\tilde{N}\) and \(I = \tilde{I}/\tilde{N}\) the proportion of the total population size remains constant. The proportion system is:

\begin{equation}\label{EQ_4_SYSTEM_2}
\begin{cases}
\dfrac{dS}{dt} = \alpha - \alpha S - \beta SI + \gamma (M) I + \nu IS\\
\dfrac{dI}{dt} = \beta SI - \left(\alpha + \nu + \gamma(M)\right) I + \nu I^{2}
\end{cases}
\end{equation}

Certainly the stationary states \(S^{*}, I^{*}\) and the basic reproductive number \(\mathcal{R}_{0} \) depend on the function \(\gamma (M) \). The reason why we chose the \(\gamma\) parameter as the only function that depends on the consumption level  \(M\) is due to the following study of sensitivity of \(\mathcal{R}_{0} \) with respect to all the other parameters shown in Tab \ref{table1}. Remember that the value of \(\mathcal{R}_{0} \) is the one of that mathematically determines when there is an epidemic. 
The basic reproductive number for the system Eq \eqref{EQ_4_SYSTEM_2} is given by
\begin{equation}\label{EQ_5_BRN}
\mathcal{R}_{0} = \frac{\beta}{\alpha + \nu + \gamma(M)}
\end{equation}

\begin{table}
\caption{Sensitivity index of \(\mathcal{R}_{0}\) (\eqref{EQ_5_BRN}) evaluated at the baseline parameter values given in Tab \eqref{table3}}
\label{table1}
\bigskip
\centering\small\setlength\tabcolsep{2pt}
\hspace*{-1cm}
\begin{tabular}{c l l r}
\hline
\textbf{Parameter} & \textbf{Sensitivity index}\\
\hline
\(\alpha\) & \(−0.0007085\) \\ 
\(\beta \) & \( + 1 \) \\
\(\gamma \) & \( - 0.998317\) \\ 
\(\nu\) & \(−0.000974275\) \\ 
\hline
\end{tabular}
\end{table}

This means that if the recovery rate increases, the value of \(\mathcal{R}_{0}\) significantly decreases.
That is to say if the level of consumption \(M\) is increased, indicating better health, the value of \(\mathcal{R}_{0}\) is expected to decrease. Regarding the death rate parameter \(\mu \), we do not study sensitivity analysis because \(\mathcal{R}_{0}\) does not depend on this parameter (for details see the electronic supplementary material).

\section{Economic Model}
In this part, an aggregate economy with a firm (business) and a representative household (individuals) will be modeled under the presence of a disease as discussed in the previous section. 
The economic model will be the Solow growth model  \cite{SOLOW} with the aggregate production function of Cobb-Douglas type,  

\begin{equation}\label{EQ_5_COBB-DOUGLAS}
\begin{split}
\tilde{Y} = c \tilde{K}^{\epsilon} \tilde{S}^{1-\epsilon} \quad \epsilon \in (0, 1)
\end{split}
\end{equation}
i.e. this economy will be closed and without government expenditure, with a homogeneous Cobb-Douglas production function  \cite{COBB,BLOOM,MOAV} of degree one where the product \(\tilde{Y}(t)\) depends on an exogenous technological parameter \(c\), of input capital \(\tilde{K}(t)\) of that period and of the population fit to work \(\tilde{S}(t)\), that in our model will be the population not infected by the disease and therefore will be the susceptible population that we modeled in the previous section, and \(\epsilon \in (0,1)\) represents the elasticity of substitution of capital in the product, that is, how important capital is in the production function.
Thus, if \(\epsilon\) takes a value very close to one, it means that the product is very capital intensive, while if \(\epsilon\) is very close to zero, the product is labor intensive. 

Using the law of motion for the stock of capital we model the dynamic of capital,

\begin{equation}\label{EQ_6_SOLOW_MODEL}
\begin{split}
\dfrac{d\tilde{K}(t)}{dt} & = \text{saving/investment} - \text{depretiation}\\
\dfrac{d\tilde{K}(t)}{dt} & = \tilde{I}_{n}(t) - \delta \tilde{K}(t)\\
\end{split}
\end{equation}
where \(\tilde{I}_{n}(t)\) is the savings/investment function. It is assumed to be of a Keynesian nature, i.e. savings (and investment in a closed economy) equals a constant fraction \(a\) of total income \(\tilde{Y}(t)\) , i.e. \(\tilde{I}_{n}(t) = a \tilde{Y}(t)\)

\begin{equation}\label{EQ_7_KEYNESIAN}
\dfrac{d\tilde{K}(t)}{dt} = a \tilde{Y}(t) - \delta \tilde{K}(t)\quad a \in [0, 1]
\end{equation}
where \(a\) represents the portion of the product to be saved and \(1-a\) the portion of the product goes to consumption.

\begin{figure}
\centering
\includegraphics[scale=0.7]{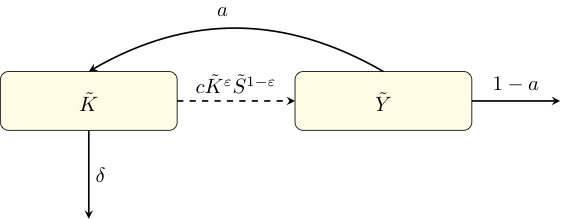}
\caption{Capital flow chart of the economic model}
\label{CAPITAL_FLOW_CHART}
\end{figure}

The susceptible population  will be now determined by both dynamics, the dynamics of disease and the dynamics of capital. The coupling functions of the two dynamic systems will be the recovery rate function \(\gamma (M)\) and the Cobb-Douglas production \(\tilde{Y}(t)\) function. This is shown in the diagram Fig \ref{CAPITAL_FLOW_CHART}, where all the savings are invested in capital for production, while capital depreciates at a rate \(\delta\).

As the population is not constant over time, the equilibrium of this economy will occur when capital and the susceptible population, at their \emph{per-capita} levels, do not change over time. With this in mind, we will use the law of movement of capital and then find the equilibrium of capital \emph{per-capita} and with this find the equilibrium of the economy. Using \(K = \tilde{K}/\tilde{N}\), \(Y = \tilde{Y}/\tilde{N}\), \(S = \tilde{S}/\tilde{N}\), \(I = \tilde{I}/\tilde{N}\) we can obtain the \emph{per-capita} capital, the \emph{per-capita} product, the susceptible and infected population in proportions and thus find an expression for the capital change. So, the dynamical system to analyze now is, 

\begin{equation}\label{EQ_9_system_3}
\begin{cases}
S\, ' = \alpha - \alpha S - \beta SI + \gamma (M) I + \nu IS\\
I\, '= \beta SI - \left(\alpha + \nu + \gamma(M)\right) I + \nu I^{2}\\
K\, '= a Y - \left(\delta + \alpha  - \mu - \nu \left(1 - S\right)\right)K\\
Y\, ' = c (1-\epsilon) K^{\epsilon}S^{-\epsilon} S\,' + 
c\, \epsilon S^{1-\epsilon}K^{\epsilon-1}K\,'
\end{cases}
\end{equation}

that has the steady states, different from the trivial equilibria,

\begin{equation}\label{EQ_9}
\begin{split}
S^{*} &= \dfrac{\alpha + \gamma(M)}{\beta - \nu}
\quad 
I^{*} = 1 - \dfrac{\alpha + \gamma(M)}{\beta - \nu} \\
K^{*} &=  \left(\dfrac{a\, c}{\delta + \alpha - \mu - \nu (1-S^{*})}\right)^{1/(1 - \epsilon)} S^{*} \\
Y^{*} &= c^{1/(1-\epsilon)} 
\left(\dfrac{a}{\delta + \alpha - \mu - \nu \left(1 - S^{*}\right)}\right)^{\epsilon/(1-\epsilon)} S^{*}
\end{split}
\end{equation}
which are stable. Recall that we also assume that a part of the product is invested in savings \(a\tilde{Y}\) and another part in consumption \(\tilde{M} = (1-a)\tilde{Y}\) or in \emph{per-capita} terms \(M = (1-a)Y\). Therefore these equilibria are chained, because \(S^{*}\) involves the recovery rate function \(\gamma(M)\) and this in turn depends on the part that will be consumed. (for more information see electronic supplementary material).

\section{Results}
We have the following analytical results produced by the coupled model of two dynamic systems, one epidemiological and another economic, which is a complex system  \cite{LEVIN}. If \(\mathcal{R}_{0}>1 \) and \(\delta \alpha > \mu + \nu\)  then the equilibria in Eq \eqref{EQ_9} are stable. (demonstration in electronic supplementary material). 

To explore the implications of our assumptions and models of the previous theoretical sections we now use numerical data. Some are found in the literature and others are averages or approximations of economies affected by recurrent diseases that satisfy our assumptions. In some cases we use data from populations that inhabit the central part of Africa, particularly Nigeria. The diseases from which we take some data and then average them are diseases that suffer from these communities. 
Some data were taken from vector-borne diseases such as dengue or malaria. Although these diseases are transmitted by vectors and from the epidemiological point of view they are modeled using other models, in our case this is not important, since we are not studying the disease as such or its transmission, we are studying individuals and their state of health and how this impacts on your physical work.

Unfortunately calculating averages is not the logical thing, because we cannot compare an individual who acquires malaria or dengue with another individual who is infected by a parasite that has produced  gastroenteritis or even with another individual who acquire two or three diseases at the same time. The question of calculating the rate of transmission when we are observing various diseases is clearly not an average. In the case of dengue, the rate of transmission is calculated as the product of daily biting rate to a person, which is related with density proportion between humans and mosquitoes by the likelihood of the bite being effective. This value cannot be averaged with the transmission rate of a parasite that produces diarrhea and is in the untreated water that people consume without any sanitary precaution such as boiling the water. In this way some values ​​ were used in the simulation without being able to give an exact reference because no results of this type of field work are found. The data and their interpretations are shown in Tab \ref{table2} and \ref{table3}.

\begin{table}
\caption{Model functions and variables}
\label{table2}
\bigskip
\centering\small\setlength\tabcolsep{2pt}
\hspace*{-1cm}
\begin{tabular}{c l l r}
\hline
\textbf{Variable} & \textbf{Interpretation}& \textbf{Units} & \\ \hline 
\(t\) & time & days & \\ 
\(M\) & \emph{per-capita} consumption  & USD & \\
\hline
\textbf{Function} & \textbf{Interpretation} & \textbf{Value} & \textbf{Cite}\\
\hline
\(S(t)\) & proportion of susceptible population & dimensionless & \\
\(I(t)\) & proportion of infected population (sick) & dimensionless & \\
\( \gamma(M) \) & recovery rate as a function of \(M\)  & \(\text{days}^{-1}\) & \\
\( \mathcal{R}_{0} (M)\) & basic reproduction number as a function of \(M\) & dimensionless  & \\
\(h(M)\) & health function (a metric to measure health level) & dimensionless &  \cite{BONDS1}\\
\(Y(t)\) & product \emph{per-capita} (real value (r.v.) of all goods produced per day) & \(\text{USD}\times \text{days}^{-1}\) & \\
\(K(t)\) & capital input \emph{per-capita} (r.v. machinery, equipment, and buildings) & \(\text{USD}\times \text{days}^{-1}\)  & \\
\(c\) & factor productivity & dimensionless & \\
\hline
\end{tabular}
\end{table}

\begin{table}
\caption{Value of parameters for simulation (Epidemic period)}
\label{table3}
\bigskip
\centering\small\setlength\tabcolsep{2pt}
\hspace*{-1cm}
\begin{tabular}{c l l r}
\hline
\textbf{Parameter} & \textbf{Interpretation} & \textbf{Value} & \textbf{Cite}\\
\hline
 & \textbf{Epidemiological parameters} & & \\ \hline 
\( LE \) & Life expectancy (Nigeria, 2013) & \(52.11\) &  \cite{WB}\\
\( F \) & Fertility rate (Nigeria, 2013) & \(3\) &  \cite{WB}\\
\( B \) & Average daily biting per day (for dengue = 2.7) & \(0.8\) &  \cite{SMITH,BECKER} \\
\( P \) & Transmission probability per bite (for dengue) & \(0.375\) &  \cite{FOCKS,MEDEIROS,BECKER} \\
\( \alpha \) & Natural rate of birth & \(F/(LE \times 365)\) & \\
\( \beta \) & Transmission rate  & \(B\times P\) &  \\
\( \mu \) & Natural rate of death & \(1/(LE \times 365)\) & \\ 
\( \nu \) & Additional death rate caused by disease & \(1/((LE - 5)\times365)\)  & \\
\hline
\( \overline{h} \) & Maximum level of nutrition attainable & \( 1\) &  \cite{BONDS1}\\
\( \overline{\gamma} \) & Rate of natural recovery without any intervention & \(1/15\) &  \cite{WHO2}\\
\( S_{0} \) & Initial susceptible population in proportion & \(80\%\) & \\
\( I_{0} \) & Initial infectious population & \(20\%\) & \\
\hline
\(c\) &  productivity factor & \(0.8352\) &  \cite{FELIPE} \\
\( \kappa \) & Half saturation constant & \(0.30\) &  \cite{BONDS1}\\
\( a \) & Proportion of product saved & \(10\%\) & \\
\( 1 - a \) & Proportion of product intended for consumption & \(90\%\) & \\
\(\epsilon\) & Elasticity of substitution of capital in the product & \(0.25\) &  \cite{FELIPE}\\
\(\delta\) & Average depreciation rate (Nigeria) of the capital stock & \(0.059\) &  \cite{KNOEMA}\\
\hline
\end{tabular}
\end{table}

\begin{figure}
	\centering
	\noindent\begin{minipage}[b]{.45\textwidth}
		\centering
		\includegraphics[width=2.2in]{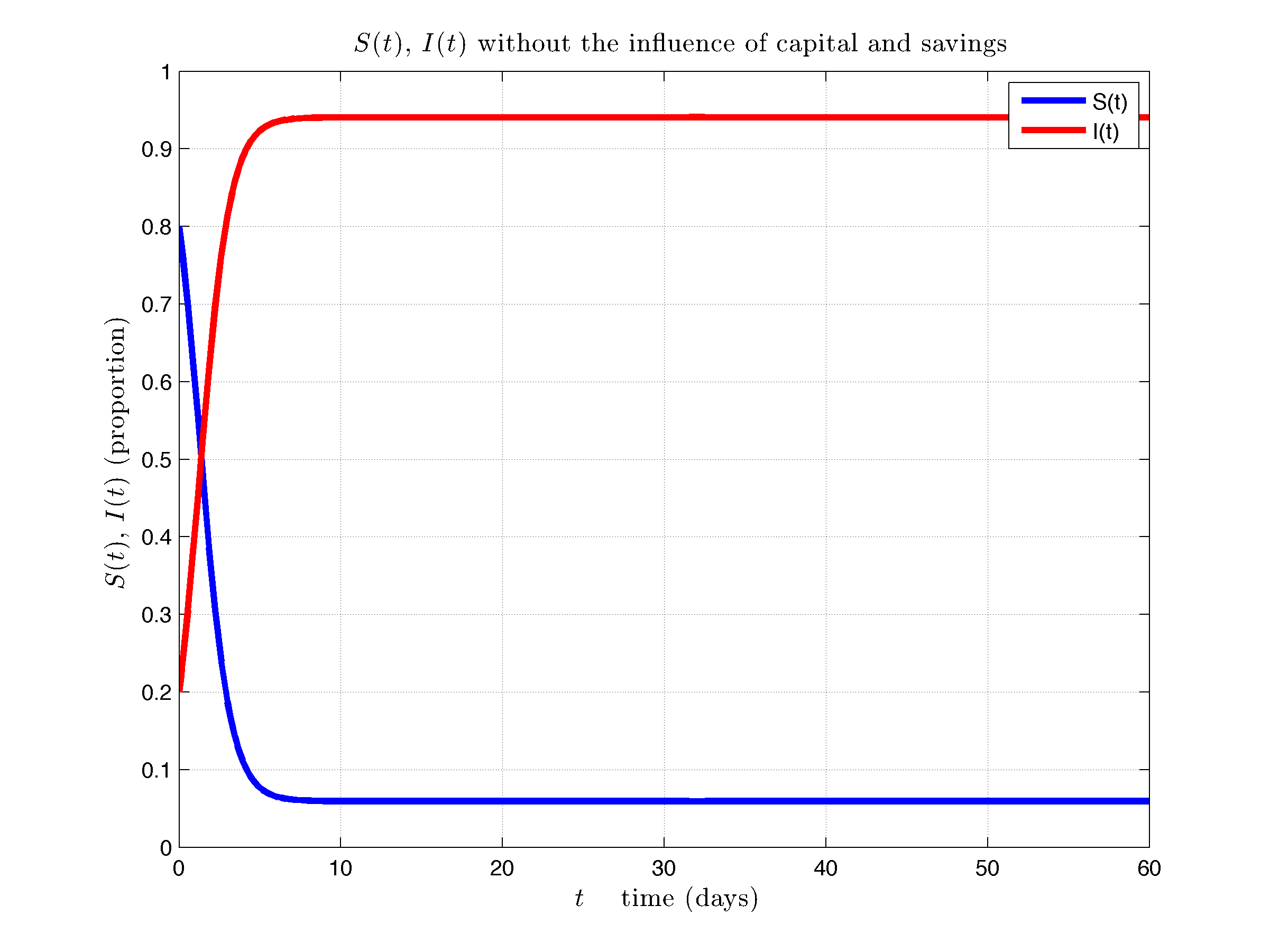}
	\end{minipage}%
	\noindent\begin{minipage}[b]{.45\textwidth}
		\centering
		\includegraphics[width=2.2in]{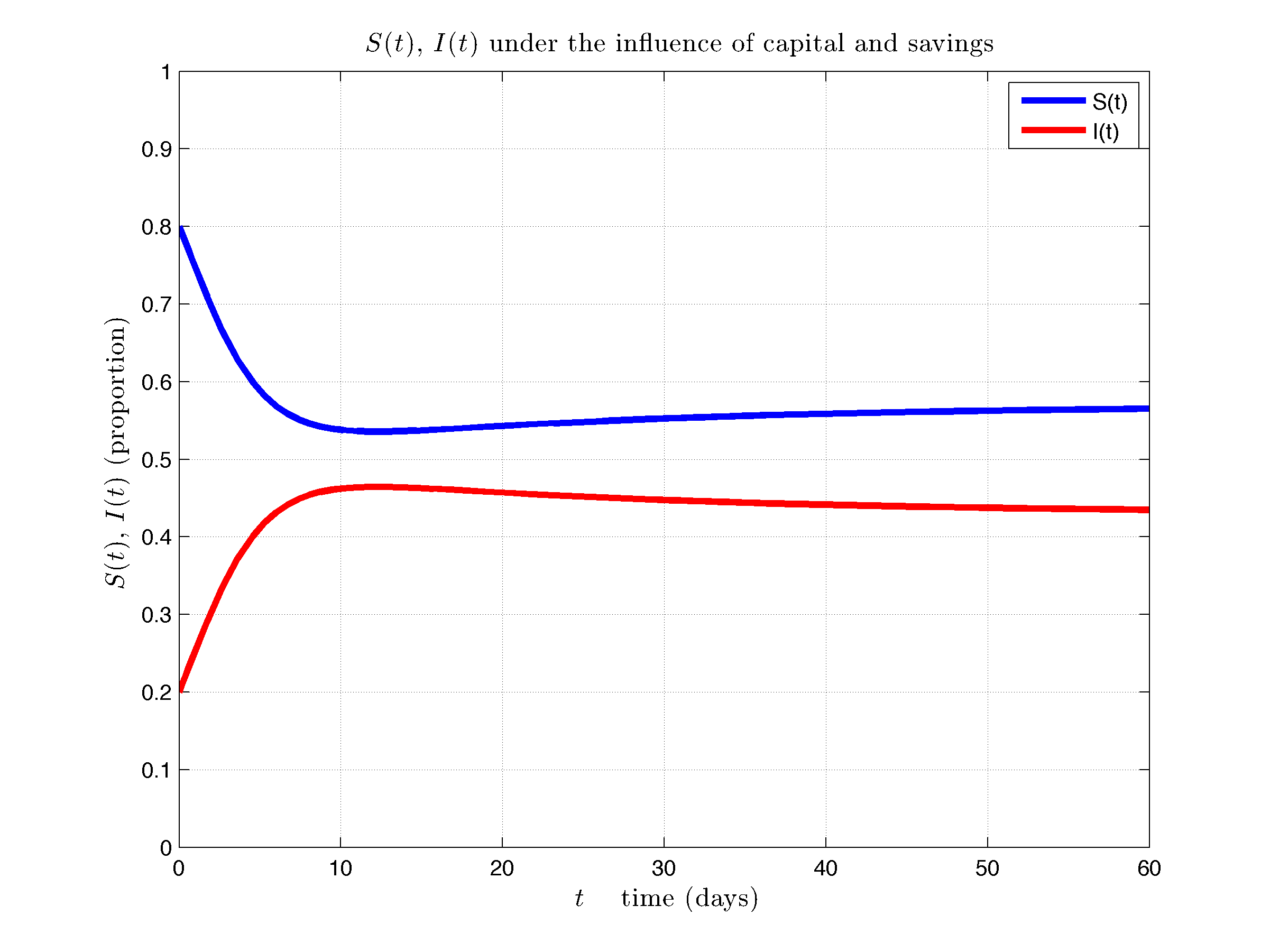}
	\end{minipage}
\caption{
The graph on the left shows the solution for the population of susceptible \(S(t)\) and infected \(I(t)\) in proportions without any influence of capital or savings (solution of the system \eqref{EQ_4_SYSTEM_2} with \(\gamma (M) = \overline{\gamma}\), 
while the graph on the right shows the population of susceptible \(S(t)\) and infected \(I(t)\) in proportions with the influence of per capita savings and of course capital (solution of the system \eqref{EQ_9_system_3})
Initial conditions are \(S(0)=0.8\), \(I(0)=0.2\), i.e. in \(t=0\) the \(80\%\) of the population is healthy, but susceptible \(S(0)\), and the infected population  \(I(0)\)  is of \(20\%\)  for \(t\) ranging from \(t = 0\) to \(t = 60\) days. The other parameters are shown in Tab \ref{table3}. The basic reproduction number \(\mathcal{R}_{0} >1\) in this epidemic period. Note the fluctuation of the populations: in the case of the susceptible individuals it decreases strongly but  has a slight recovery, whereas the population of infected increases rapidly but soon it diminishes slightly.}
\label{SOL_SI}
\end{figure}

Fig \ref{SOL_SI} shows the solution for the population of susceptible \(S(t)\) and infected \(I(t)\) in proportions without any influence of capital or savings (solution of the system \eqref{EQ_4_SYSTEM_2} with \(\gamma (M) = \overline{\gamma}\), 
and the same populations with the influence of per capita savings and of course capital (solution of the system \eqref{EQ_9_system_3})
Note the fluctuation of the populations, in the case of the susceptible individuals it decreases strongly but  has a slight recovery, whereas the population of infected individuals increases rapidly but soon it diminishes slightly. The initial values ​​in this simulation for the solution of the system \eqref{EQ_9_system_3} are: \(S(0) = 0.8\), \(I(0) = 0.2\), \(K(0) = 0.3\) and using Cobb-Douglas formula \(Y(0) = 0.522863457\).  
Note the fluctuation of the populations in Fig \ref{SOL_SI} which shows the solutions for the susceptible and infected populations of the systems \eqref{EQ_4_SYSTEM_2} and \eqref{EQ_9_system_3}. Note that when there is no capital inflow and saving, the susceptible population  decreases to a stable steady state and the infected population grows to a steady state (left side graph). However, when there is capital inflow and saving the population of susceptible individuals decreases, reaches a minimum and then grows until reaching its stable equilibrium point. For this same case the population of infected grows, reaches a maximum and then decreases until reaching its stable steady state (right side graph). That is, the model is showing recovery of the susceptible population in an epidemic due to the influence of the permanent saving of the susceptible population who are the ones that produce directly.

\begin{figure}
\centering
\noindent\begin{minipage}[b]{.45\textwidth}
\centering
\includegraphics[width=2.2in]{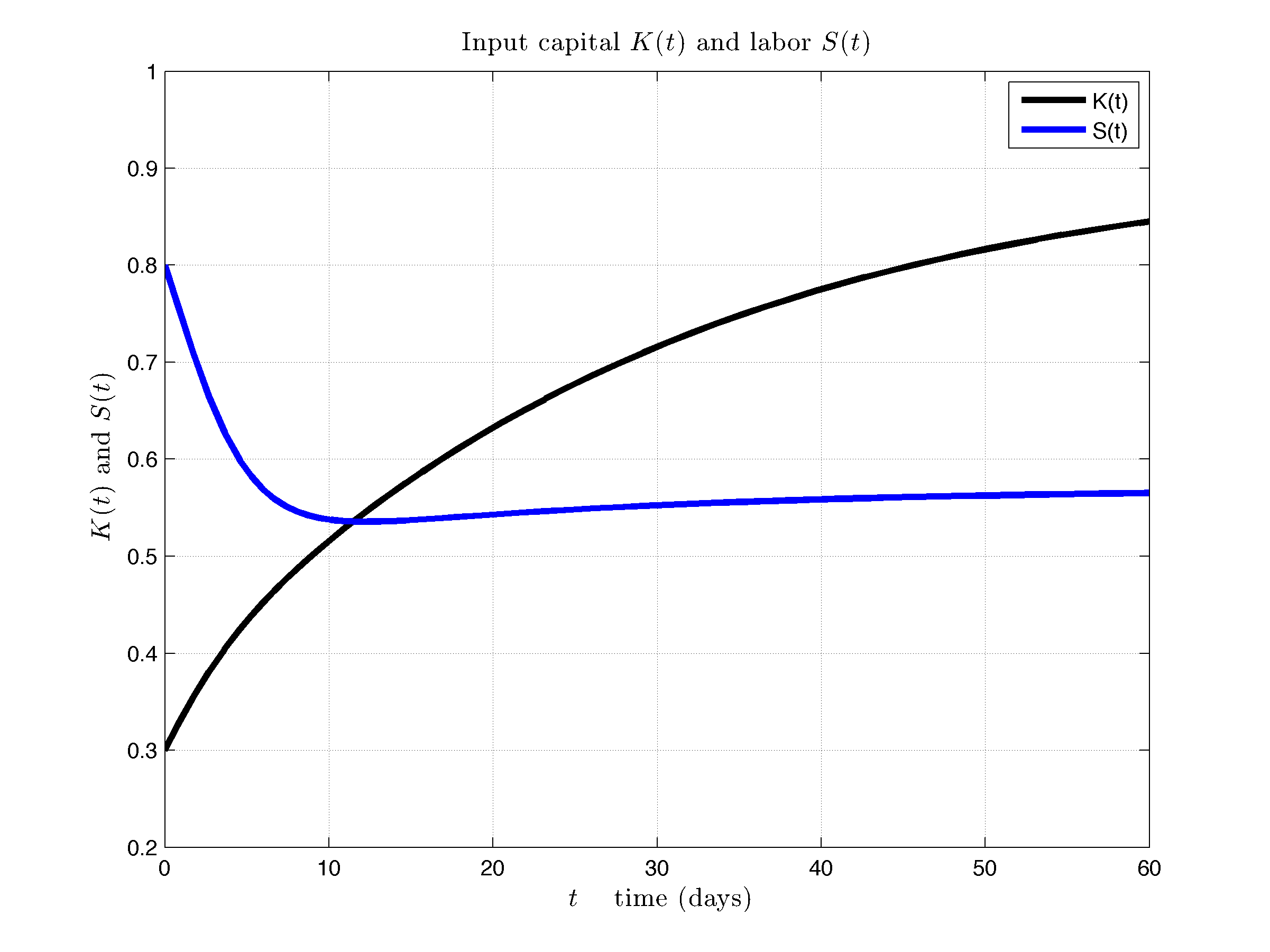}
\end{minipage}%
\noindent\begin{minipage}[b]{.45\textwidth}
\centering
\includegraphics[width=2.2in]{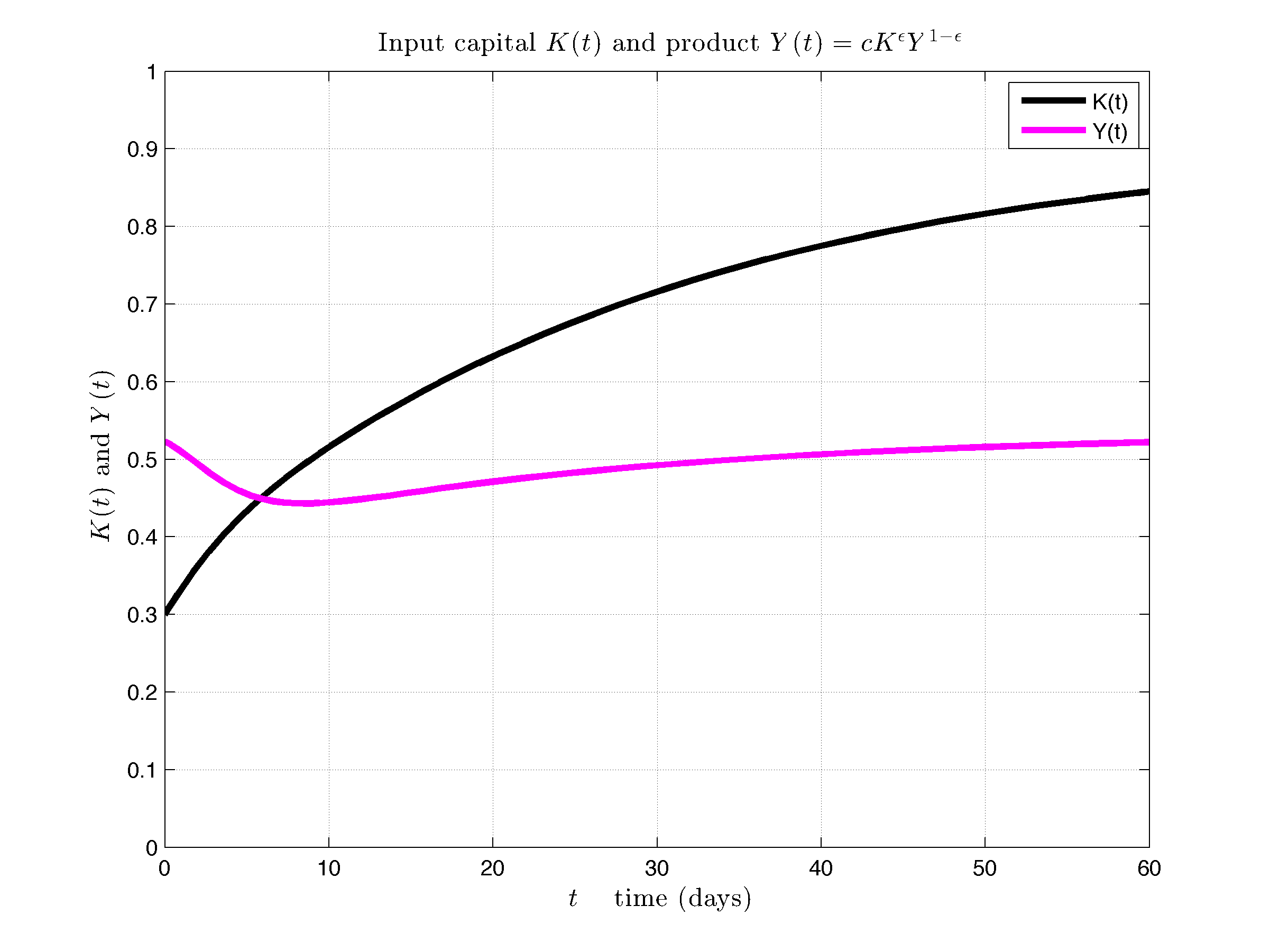}
\end{minipage}
\caption{
The graph on the left shows the solution for input capital \emph{per capita} \(K(t)\) and susceptible population \(S(t)\) in proportions (solution of the system \eqref{EQ_9_system_3}). 
Initial conditions are \(S(0)=0.8\), \(K(0)=0.3\) ranging from \(t = 0\) to \(t = 60\) days. The other parameters are shown in Tab \ref{table3}. The basic reproduction number \(\mathcal{R}_{0} >1\) in this epidemic period. Note the following surprising result of the model:  While capital grows and tends to stabilize the product decreases in the presence of the epidemic but it manages to recover until reaching its level of stable equilibrium which is a little more than the initial value.}
\label{SOL_KS_KY}
\end{figure}

Fig \ref{SOL_KS_KY}  shows the other solutions of system \eqref{EQ_9_system_3}: the behavior of capital and the production of an economy affected by an epidemic. 
Note that capital is always increasing  due to the permanent saving of the susceptible population. Note also that the susceptible population  decreases only at the beginning reaching a minimum value from which it begins to grow and reaches its stable steady state. 
This same behavior happens to the economic production. 
This effect shows the importance of saving the susceptible population which is the only one that contributes to the economy its work force and its capacity of saving resulting in the recovery of the susceptible population and therefore of the economic production.

What is surprising and not expected is that the model shows that in this case the production initially decreases but then recovers until it reaches initial production levels. Moreover the point of equilibrium manages to slightly exceed the initial level of production.

\begin{figure}
\centering
\noindent
\begin{minipage}[b]{.45\textwidth}
\centering
\includegraphics[width=2.2in]{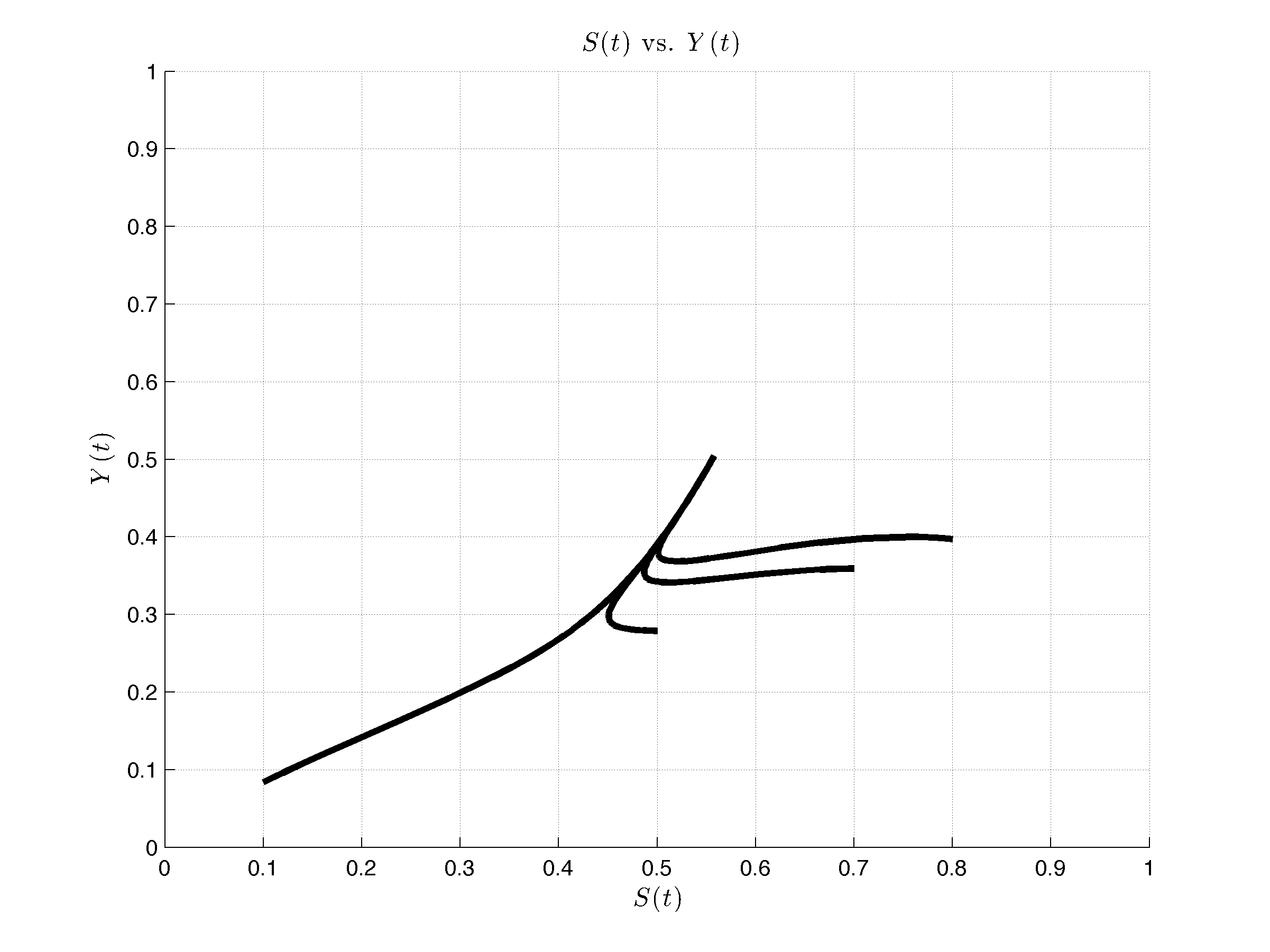}
\caption*{\(a = 0.1\) and \(K(0) = 0.1\)}
\end{minipage}%
\noindent\begin{minipage}[b]{.45\textwidth}
\centering
\includegraphics[width=2.2in]{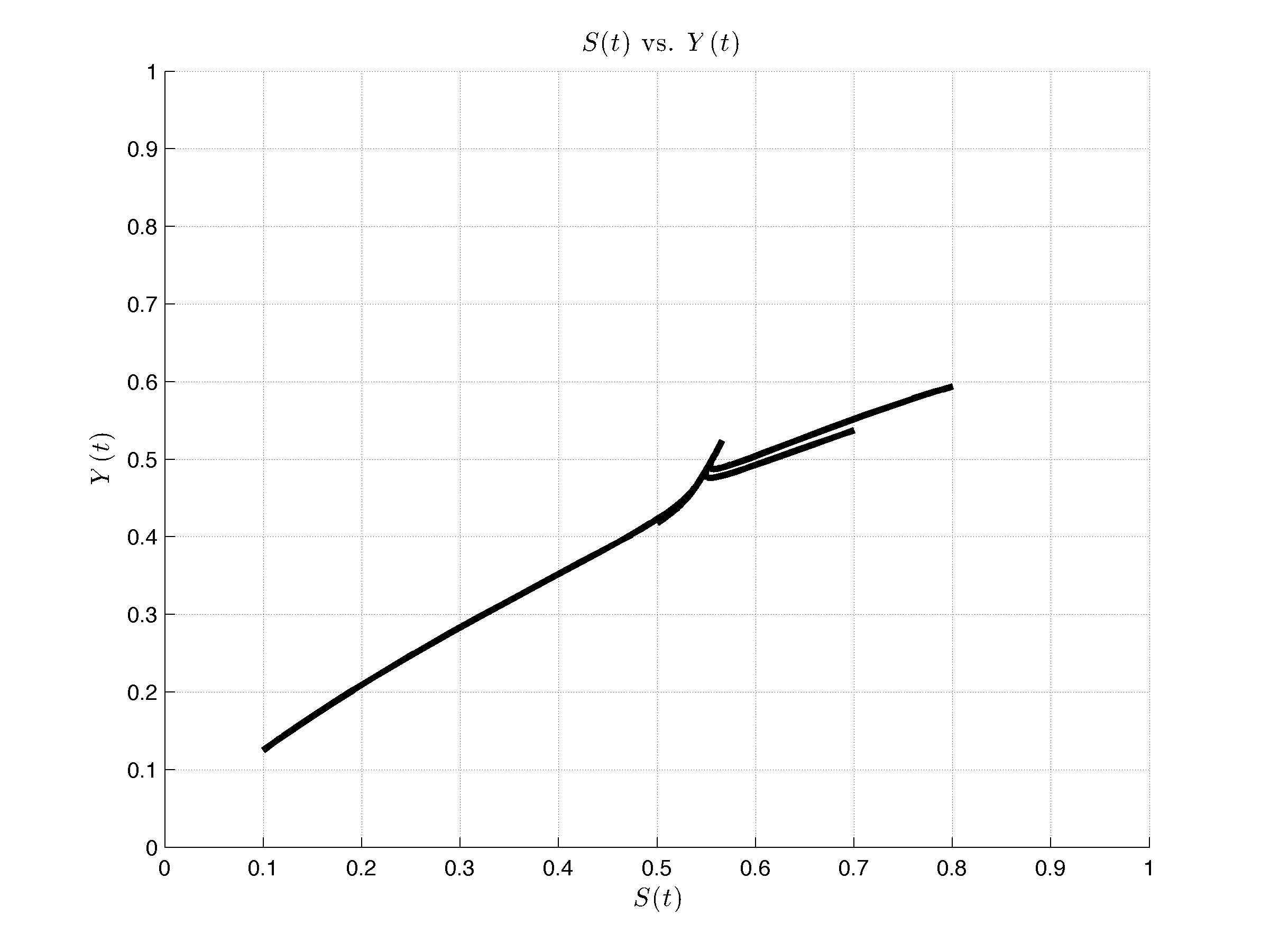}
\caption*{\(a = 0.1\) and \(K(0) = 0.5\)}
\end{minipage}
\\
\begin{minipage}[b]{.45\textwidth}
\centering
\includegraphics[width=2.2in]{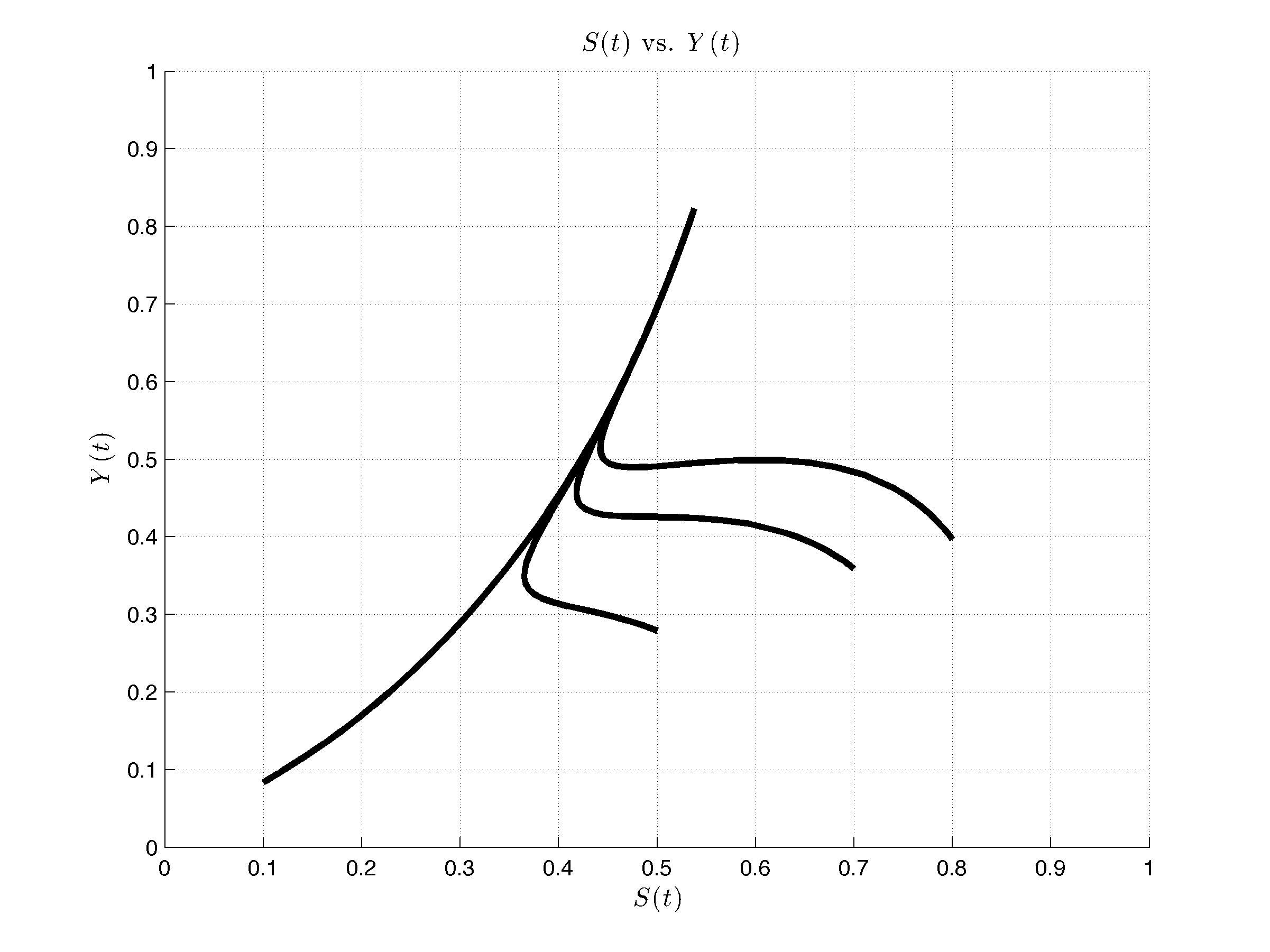}
\caption*{\(a = 0.5\) and \(K(0) = 0.1\)}
\end{minipage}%
\noindent\begin{minipage}[b]{.45\textwidth}
\centering
\includegraphics[width=2.2in]{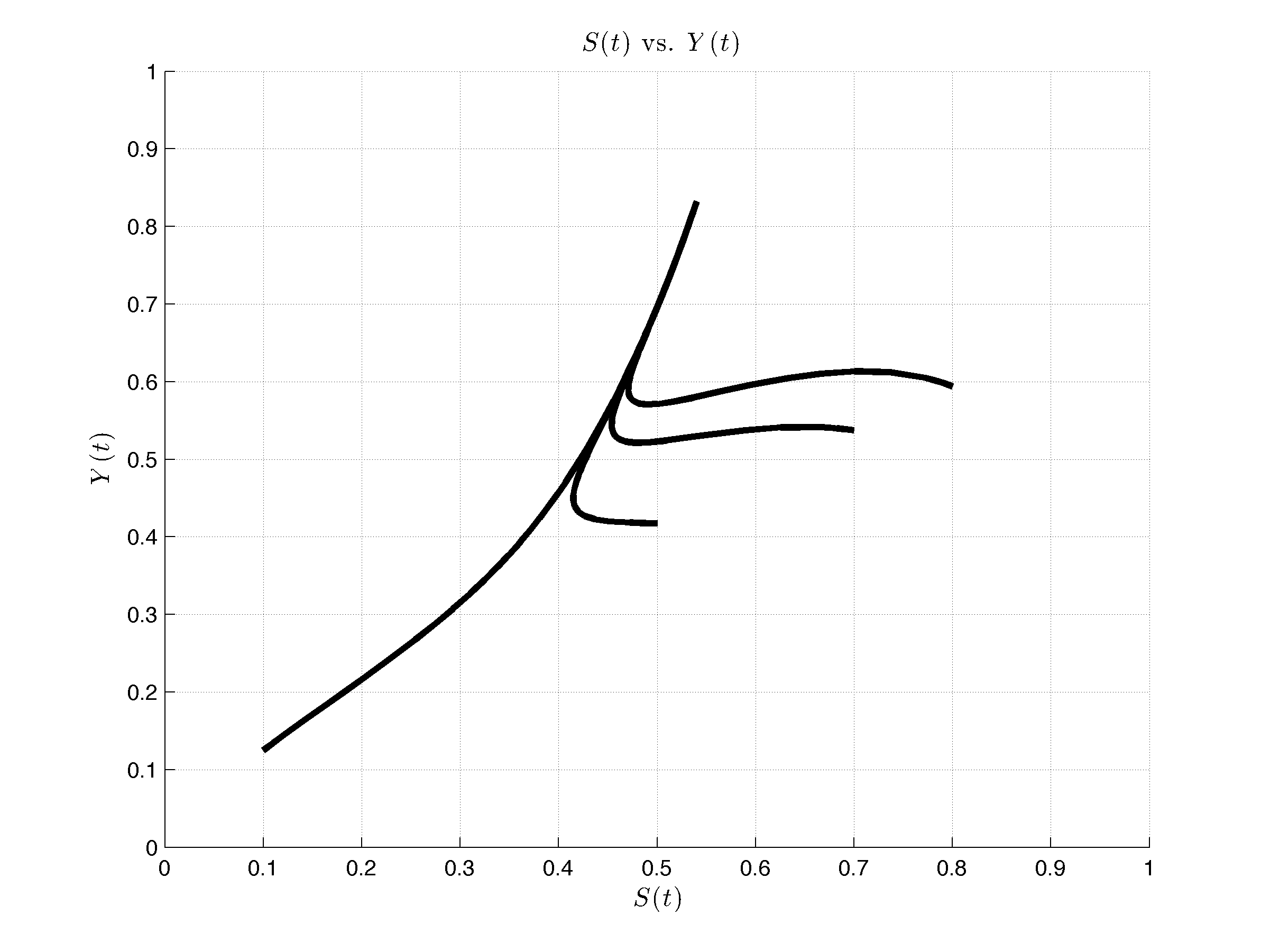}
\caption*{\(a = 0.5\) and \(K(0) = 0.5\)}
\end{minipage}
\caption{Phase portrait of \emph{per-capita} product, \(Y(t)\) (on the vertical axis), versus susceptible population in proportions, \(S(t)\) (on the horizontal axis).
In all four graphs the initial value \(S (0) \in \left\{0.1, 0.5, 0.7, 0.8 \right\}\).
For the two top charts the \emph{per capita} saving level is equal to \(10\%\), (\(a=0.1\)), while for the two below the saving level is \(50\%\), (\(a=0.5\)). 
The initial capital, for the two graphs on the left side, is \(10\%\), (\( K(0) = 0.1\)), while for the two on the right side is \(50\%\), (\(K(0) = 0.5\)). 
The value corresponding to \(Y(0)\) is obtained by the Cobb-Douglas formula \eqref{EQ_5_COBB-DOUGLAS}.
Under epidemic conditions, ie under the assumption that the epidemic does not end, the levels of recovery of the susceptible population and of production depend more heavily on saving (\(a\)) than on initial capital introduced into the economy (\(K(0)\)). Note that with an initial capital of \(50\%\) the recovery of these two variables depends on the saving level. Thus, in communities that can not have a medium or high initial capital, ie communities that can only have low initial capital \emph{per capita}, the susceptible population and the production recover only if there is permanent saving. Although it is very low what is important is the savings.
The basic reproductive number is \(\mathcal{R}_0 > 1 \). The parameters are in Tab \ref{table3} and the time is \(t = 0, \dots 1500 \) days.
}
\label{PP_SY}
\end{figure}

Fig \ref{PP_SY} shows  the phase portrait of production level \(Y(t)\) vs susceptible population \(S(t)\). 
The objective of this simulation of our model is to highlight the recovery of both the susceptible population and the production level  in an economy affected by an epidemic under the assumptions that the susceptible population is the only population that produces and a portion of its income is saved and that these saving is reinvested back into capital. 
It is expected that the susceptible population and therefore the production level will decrease and effectively the model shows it in the first days of the epidemic (\(10\) first days for this simulation) but, from this moment there are a recovery of the population and the production level until reaching their equilibrium. 
This is the most important result of the model and this happens due to two things: first the permanent saving and second to the reinvestment of this saving to capital. 

We can see clearly where these minimum points are from which the susceptible population and the economy recover. The important thing that is shown is the recovery of the economy affected by an epidemic only assuming that the population saves permanently. The susceptible population also recovers and then reaches equilibrium.
The difference between the four simulation charts in Fig \ref{PP_SY} is only in the percentage saved and the initial capital percentage.
Note that under epidemic conditions the  recovery level of the susceptible population and of production depend more heavily on saving (\(a\)) than on input capital (\(K(0)\)). Note that with an initial capital of \(50\%\) the recovery of these two variables depends strongly on the saving level. Thus, in poor communities that can only have low initial \emph{per capita} capital, the susceptible population and the production recover only if there is permanent saving. The conclusion that follows from the model is the importance of saving.

\begin{figure}
\centering
\includegraphics[scale=0.4]{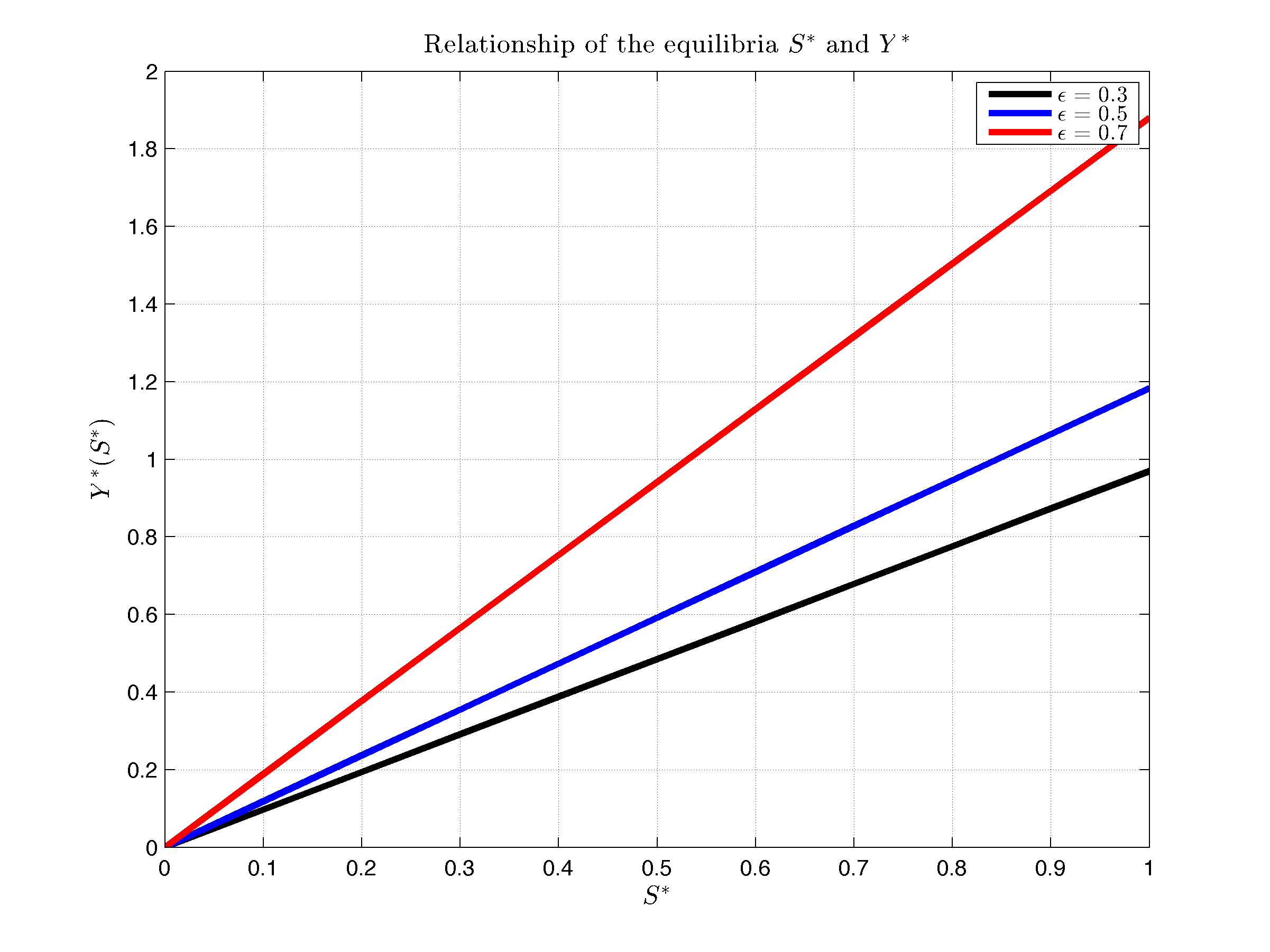}
\caption{The model shows an expected result, the higher the elasticity of input capital, the greater the output product. In this case  we are increasing  \(\epsilon\): \(0.3\) \(0.5\) \(0.7\). }
\label{SY_EPSILON}
\end{figure}

We can see the effect that it has on the equilibria by varying the values ​​of the elasticity product of  input capital \(\epsilon\).  
Fig \ref{SY_EPSILON} shows us the effect of increasing the elasticity of input capital \(\epsilon\), i.e. buying machinery, buildings, etc., on the equilibrium of the economy.

Note that the equilibrium of the product as a function of the equilibrium of the susceptible population is almost linear and it grows as we increase the product elasticity of capital \(\epsilon\).
This is one of the expected results of the model. 
However,  Fig \ref{SY_EPSILON} is not capturing the full effect of the proportion \(S^{*}\) on \(Y^{*}\), as long as the assumption is made that the saving rate is a parameter of the model and does not capture the trade-off that the home must make for its decision to save to have more resources in the next period, or consume in this period to stay healthy and to be able to produce. 
It is in this sense that there is an implicit decision of households in the proportion of the \emph{per capita} product they save and the proportion they consume. 

\begin{figure}
\centering
\includegraphics[scale=0.6]{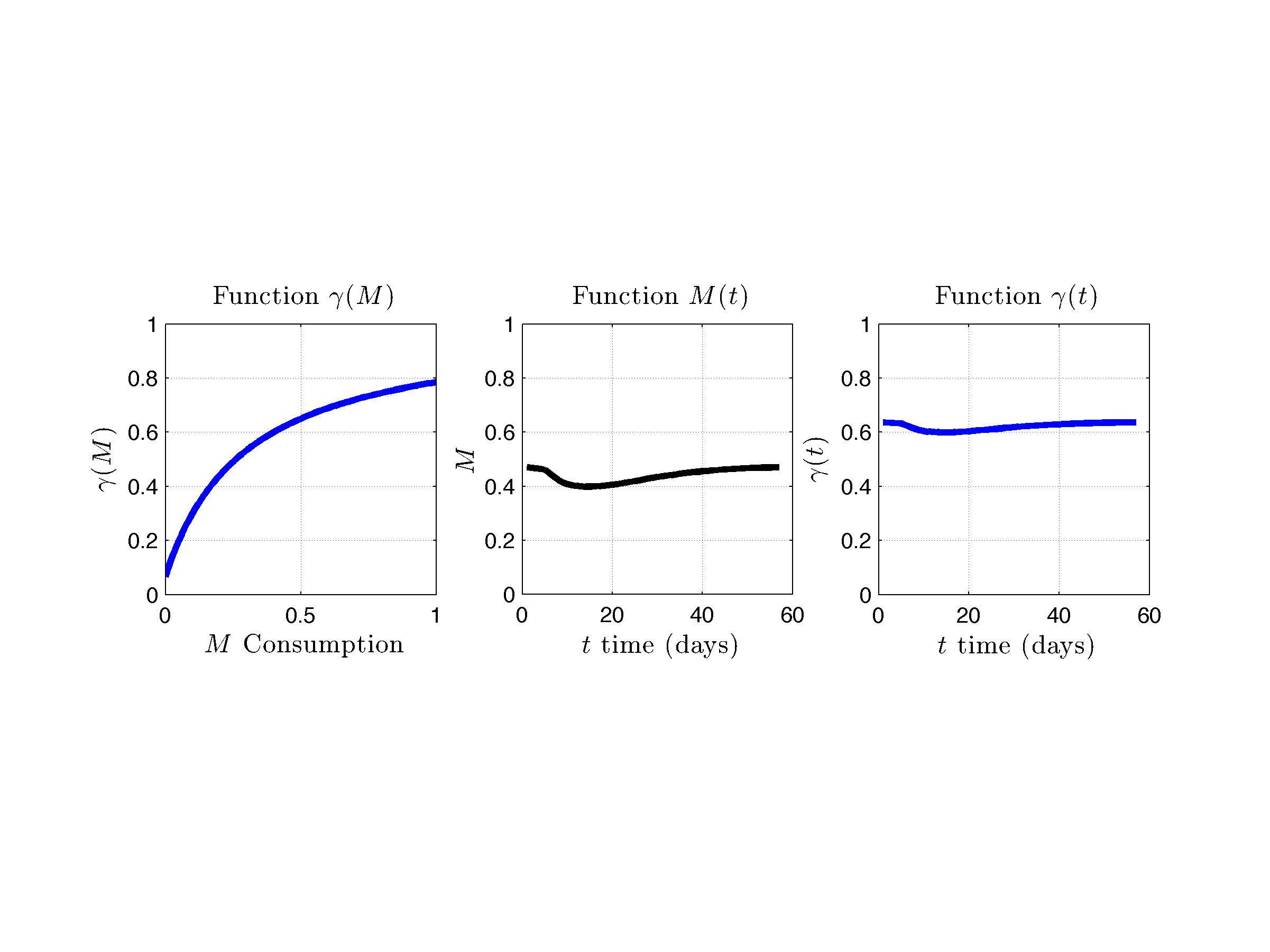}
\caption{To show that the recovery parameter \(\gamma\) is a function on the time \(t\) note the following sequence: by definition \(\gamma = \gamma (M) \) and the consumption \( M = M (a, Y) \) but \(Y = Y (t)\). Thus \(M = M(t)\) and it has an equilibrium. 
Fig A on the left shows the recovery parameter as a function of consumption \(\gamma = \gamma(M) \). Fig B (center) shows the consumption as a function of time \(M = M (t)\). Fig C on the right shows the recovery parameter as a function of time \(\gamma = \gamma (t) \).
}
\label{GAMMA}
\end{figure}

The variables susceptible population \(S\), infected population \(I\), input capital \(K\) and production \(Y\) depend on time \(t\) (see Eq \eqref{EQ_9_system_3}). 
To understand that the disease recovery parameter \(\gamma\) also depends on the time \(t\) we can follow the following dependency sequence (see Fig \ref{GAMMA}): the recovery parameter \(\gamma\) is defined as a function of the income \(M\), \(\gamma = \gamma (M)\) (see Fig A). 
In turn consumption is defined as a part of production (Solow model for a closed economy), \(M = (1-a) Y \) and the other part of production was saved \(\text{Savings} = a Y \) where \(a\) is the percentage of production that is saved. The remaining part is reinvested in capital. 
Therefore consumption depends on time (see Fig B). 
In this way we can observe that the recovery parameter \(\ gamma\) also depends on the time \(\gamma = \gamma (t)\) (Fig C). 
Thus the parameter \(\gamma\) depends on the time \(t\) and also has its stable equilibrium. 
The equilibria  \(S^{*}\), \(I^{*}\), \(K^{*}\) and \(Y^{*}\) will depend only on the percentage saved \(a\) and on the other fixed parameters.

Eq \eqref{EQ_10} shows  the \emph{per capita} product at equilibrium, \(Y^{*}\), as a function of the savings ratio of the economy, \(a\), where \(\gamma(a)\) represents the recovery rate, which as previously explained, would depend on the consumption and therefore on the savings rate.

\begin{equation}\label{EQ_10}
\begin{split}
\gamma (a) &=  \dfrac{(1 - a) Y \overline{h} + \kappa \overline{\gamma} }{(1-a)Y + \kappa} \\
S^{*}(a) &= \dfrac{\alpha + \gamma (a)}{\beta - \nu}\\
Y^{*}(a) &= c^{1/(1-\epsilon)} 
\left(\dfrac{a}{\delta + \alpha - \mu -  \nu \left(1 - S^{*}(a)\right)}\right)^{\epsilon/(1 - \epsilon)} S^{*}(a)
\end{split}
\end{equation}

Therefore, we can obtain an expression for the equilibrium of the product in terms of the savings percentage of the economy. In other words, we can find a function of product equilibrium in terms of saving , \(Y^{*} = Y^{*}(a)\). With this function we can find the level of savings that maximizes the equilibrium of the economy's \emph{per capita} product.

\begin{figure}
	\centering
	\includegraphics[scale=0.4]{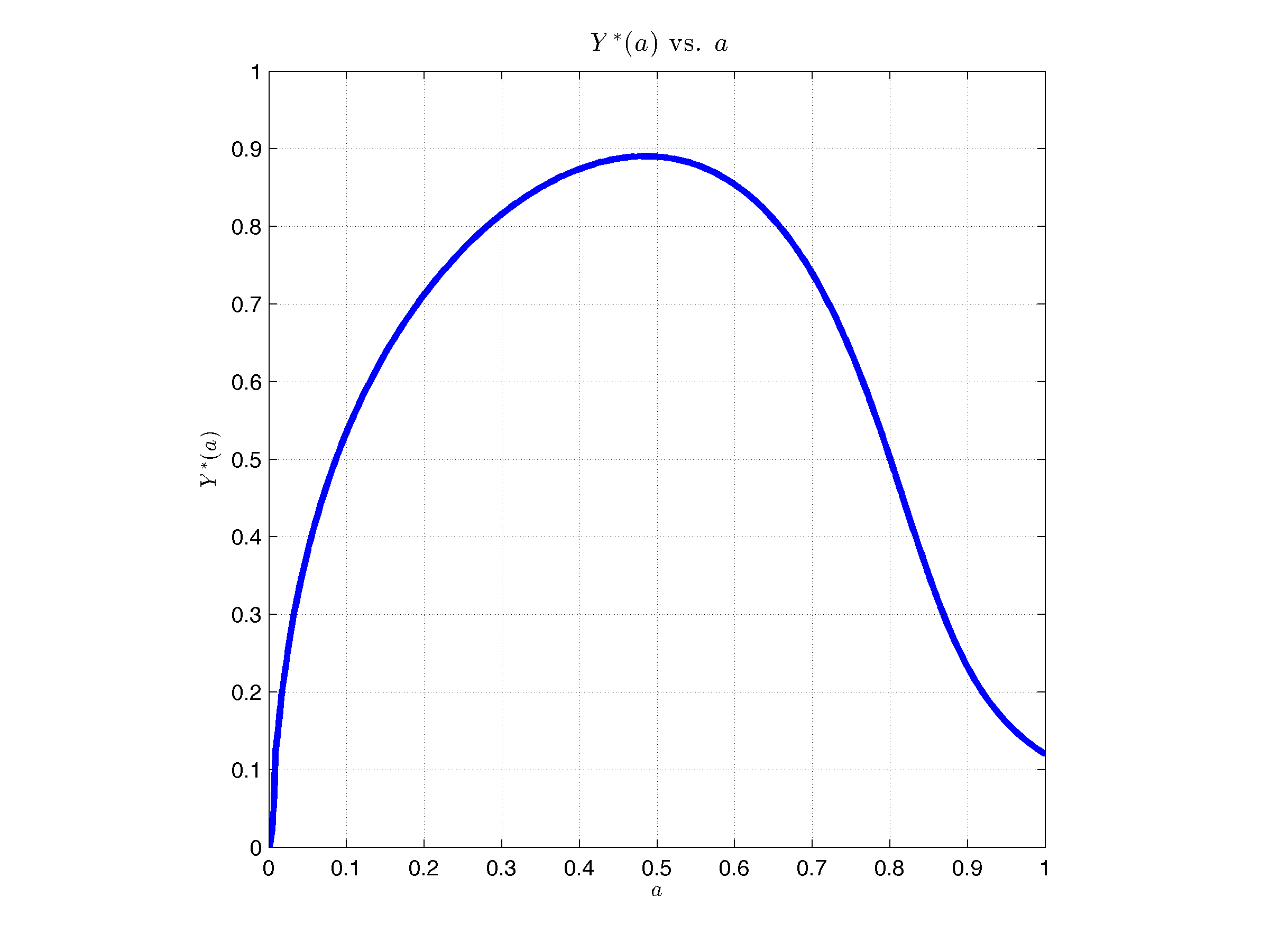}
	\caption{The model shows a surprising result: the optimum level of savings \(a\) that maximizes the product \(Y(a)\) is approximately \(50\%\) of the income.  Note that with only \(10\%\) of the savings reach the half of the product is possible.}
	\label{SY_SAVING}
\end{figure}

Fig \ref{SY_SAVING} shows the result that answers our initial question:  the dependence of the product of an aggregate economy under the influence of a pandemic disease and the percentage of \emph{per capita} saving that achieves a maximum product. 
This is perhaps the most important result  within the whole model of coupled epidemiological and economic dynamics. 
This is not an expected result, since what is expected by simple intuition, in a community with a poor economy and also affected by an endemic disease or set of diseases is to save absolutely nothing. 
But what the model is showing us is that saving is good because the product is growing until a typical individual of this community  saves half of its incomes.
The optimum level of savings to maximize the product is close to \(50\%\). 
Note that the level of production is the same saving \(20\%\)  as \(70\%\). (for more information see electronic supplementary material).

\section{Discussion}

In  Eq \eqref{EQ_9} and \eqref{EQ_10} we find the equilibrium of the product of the economy that depends only on the model parameters. This implies that the behavior of the economy in the long term
depends solely on economic variables such as: production technology, the elasticity of substitution of capital, savings and the rate of depreciation of capital and  on  epidemiological variables such as: birth rate, mortality rate (natural and due to disease), and the rate of the infected population.

At first glance, the model presented would seem to replicate the economic effects that have been studied and developed in the classical economic literature, where savings and production technology  positively affect the product, while the depreciation and level of consumption negatively affect the product. 
However, our model captures an additional effect:  capital investment in health through consumption.

In this sense, the effect of saving on \emph{per capita} production is not altogether trivial, since although a higher level of savings implies an increase of the per capita product, in turn it implies a decrease in the proportion of susceptible population, degrading the health of the economy and therefore reducing the susceptible population, which represents the basis of labor.
Even worse, as can be seen in the Eq \eqref{EQ_10} and Fig \ref{SY_SAVING}, at this point it is still unclear what effect the proportion of susceptible population and \emph{per capita} product has.
Although there is a direct positive effect of the susceptible population on the per capita product, the susceptible population has an indirect negative effect. As we can see there is a very strong dependence between the points of equilibrium and the level of savings. 

The economic interpretation of this phenomenon is supremely interesting, as the model would be suggesting a Malthusian and scandalous conclusion in which in this economy, under certain conditions, it is preferable not to invest in decreasing the proportion of infected, but it is desirable that the
proportion of infected increases in order to increase the mortality rate and thus, those who survive have a higher level of output \emph{per capita}. To better understand this statement, we can see the graph of \(Y^{*}\) versus \(S^{*}\) Fig \ref{SY_EPSILON}, in which the equilibrium \emph{per capita} product is plotted as a function of the proportion of susceptible population, for 3 economies with different levels of elasticity of capital substitution. 
In this graph, we can see that increasing the elasticity of capital substitution also increases  the marginal productivity of the susceptible population and to this extent, for values of \(\epsilon\) very close to 1, increasing \(S^{*}\) is not going to translate into a significant increase in output \emph{per capita}, but if will be very expensive, while increasing the proportion of susceptible consumption, would decreases savings and investment, and therefore capital and product \emph{per capita}.

\section{Conclusions}

This paper couples two very simple models of two very different branches of knowledge. 
Perhaps this is the greatest virtue and weakness of the model: the strong assumptions make the models are the simplest and this can be a limitation of what happens in reality. But in turn this simple model captures and reproduces very complex dynamics, which are the key to understanding the notions of the effect of an epidemic on an economy. This is a phenomenon that has not studied, at least quantitatively.

Regarding the epidemiological model, it is important to mention that since it is a SIS model, it turns out to be very restrictive in terms of the diseases it can model, which can have negative repercussions on the overall importance of its development. 
But since we are not studying the effects of a particular disease on an economy we have taken a nonexistent, unnamed, disease, which in the place of being an existing specific disease is a set of existing diseases that affect the dynamics of an economy. For this reason, despite the strongly assumptions that finally led us to take the simpler models the results of the coupled model are excellent.

An interesting conclusion of this work is the interdependence between health and the product of an economy. In the literature, many papers concentrate on a unidirectional effect from one element to another, but from the integration of these two models it is evident that this is a complex relationship where the effects can flow in both directions.
To understand and quantify this relationship has important benefits for public policies, since health expenditures compete for public resources it is important to have in perspective their contribution to development in order to make better informed investment decisions.

The most important result that answers our question: Why is it important to save? Is to have been able to find a function that quantifies the relationship between savings and product. The product reaches its maximum when saving is \(50\%\) of \emph{per capita} income. What is surprising and not expected is that with only saving   \(10\%\) is reached the half of maximum product.

Another of the most important conclusions reached by this work is the existence of a Malthusian poverty trap in economies with low levels of capital investment that feature pandemic disease, where the optimal decision of a central planner in these economies
is to allow the death of a large part of the population so that the economy has an optimal growth and, to this extent, population growth will be stagnant at an optimal value depending on the conditions of the economy and disease. 
Intervening in this type of economy is extremely complex, because although the increase in the mortality rate is evidently
undesirable for the quality of life of the population, the increase in life expectancy implies a deterioration in the quality of life of all, while the same resources (or even less) are being
distributed among more people. 
Thus, from the outset, the problem of policy-making in this type of economy must have a value judgment and a utilitarian or humanitarian social utility function (Rawlsian).

Another important result in this work is the existence of an optimal level of savings that maximizes the per capita product of the economy, which is again Machiavellian, but has important implications for market efficiency: maximizing individual well-being, can achieve a welfare optimum social.
This happen1s as long as the product \emph{per capita} is taken as a measure of individual well-being.

This project was supported by the Departamento de Matemáticas and the Facultad de Ciencias at Universidad de los Andes.


\bibliographystyle{elsarticle-num-names}
\bibliography{savings}

\begin{thebibliography}{28}
\providecommand{\natexlab}[1]{#1}
\providecommand{\url}[1]{\texttt{#1}}
\providecommand{\urlprefix}{URL }
\expandafter\ifx\csname urlstyle\endcsname\relax
  \providecommand{\doi}[1]{doi:\discretionary{}{}{}#1}\else
  \providecommand{\doi}[1]{doi:\discretionary{}{}{}\begingroup
  \urlstyle{rm}\url{#1}\endgroup}\fi
\providecommand{\bibinfo}[2]{#2}

\bibitem[{Reinert(2007)}]{REINERT}
\bibinfo{author}{E.~Reinert}, \bibinfo{title}{How Rich Countries Got Rich and
  Why Poor Countries Stay Poor}, \bibinfo{publisher}{New York: Carroll \&
  Graf}, ISBN \bibinfo{isbn}{9780786718429}, \bibinfo{year}{2007}.

\bibitem[{WHO(2007)}]{WHO1}
\bibinfo{author}{WHO}, \bibinfo{title}{World Health Report 2004. Changing
  history}, \bibinfo{journal}{http://www.who.int/whr/2004/en/} .

\bibitem[{WHO(2010)}]{WHO2}
\bibinfo{author}{WHO}, \bibinfo{title}{Quantifying environmental health
  impacts. Global Burden of Disease (GBD)},
  \bibinfo{journal}{http://www.who.int/quantifying
  ehimpacts\-/global/ebdcountgroup/en/} .

\bibitem[{Schmidt et~al.(2011)Schmidt, Genser, Luby, and Chalabi}]{SCHMIDT}
\bibinfo{author}{W.~Schmidt}, \bibinfo{author}{B.~Genser},
  \bibinfo{author}{S.~Luby}, \bibinfo{author}{Z.~Chalabi},
  \bibinfo{title}{Estimating the effect of recurrent infectious diseases on
  nutritional status: sampling frequency, sample-size, and bias},
  \bibinfo{journal}{J Health Popul Nutr}
  \bibinfo{volume}{29}~(\bibinfo{number}{4}) (\bibinfo{year}{2011})
  \bibinfo{pages}{317--326}, \doi{\bibinfo{doi}{10.1371/journal.pbio.1001827}}.

\bibitem[{Dillon et~al.(2014)Dillon, Friedman, and Serneels}]{DILLON}
\bibinfo{author}{A.~Dillon}, \bibinfo{author}{J.~Friedman},
  \bibinfo{author}{P.~Serneels}, \bibinfo{title}{Health Information, Treatment,
  and Worker Productivity Experimental Evidence from Malaria Testing and
  Treatment among Nigerian Sugarcane Cutters}, \bibinfo{journal}{Development
  Research Group Poverty and Inequality Team}
  \bibinfo{volume}{1}~(\bibinfo{number}{7120}) (\bibinfo{year}{2014})
  \bibinfo{pages}{1--46}, \doi{\bibinfo{doi}{10.1596/1813-9450-7120}}.

\bibitem[{Sachs and Malaney(2002)}]{SACHS}
\bibinfo{author}{J.~Sachs}, \bibinfo{author}{P.~Malaney}, \bibinfo{title}{The
  economic and social burden of malaria}, \bibinfo{journal}{Nature}
  \bibinfo{volume}{415} (\bibinfo{year}{2002}) \bibinfo{pages}{680--685},
  \doi{\bibinfo{doi}{10.1038/415680a}}.

\bibitem[{Guy(2017)}]{GUY}
\bibinfo{author}{F.~Guy}, \bibinfo{title}{Saving from poverty: A critical
  review of Individual Development Accounts}, \bibinfo{journal}{Critical Social
  Policy SAGE} \bibinfo{volume}{37}~(\bibinfo{number}{3})
  (\bibinfo{year}{2017}) \bibinfo{pages}{1--20},
  \doi{\bibinfo{doi}{10.1177/0261018317695451}}.

\bibitem[{Bonds et~al.(2009)Bonds, Keenan, Rohani, and Sachs}]{BONDS1}
\bibinfo{author}{M.~Bonds}, \bibinfo{author}{D.~Keenan},
  \bibinfo{author}{P.~Rohani}, \bibinfo{author}{J.~Sachs},
  \bibinfo{title}{Poverty trap formed by ecology of infectious diseases},
  \bibinfo{journal}{Proc. R. Soc. B}
  \bibinfo{volume}{283}~(\bibinfo{number}{1827}) (\bibinfo{year}{2009})
  \bibinfo{pages}{1185--1192}, \doi{\bibinfo{doi}{10.1098/rspb.2009.1778}}.

\bibitem[{Ngonghala et~al.(2014)Ngonghala, Plucin, Murray, Farmer, Barrett,
  Keenan, and Bonds}]{BONDS2}
\bibinfo{author}{C.~Ngonghala}, \bibinfo{author}{M.~Plucin},
  \bibinfo{author}{M.~Murray}, \bibinfo{author}{P.~Farmer},
  \bibinfo{author}{C.~Barrett}, \bibinfo{author}{D.~Keenan},
  \bibinfo{author}{M.~Bonds}, \bibinfo{title}{Poverty, Disease, and the Ecology
  of Complex Systems}, \bibinfo{journal}{PLoS Biol}
  \bibinfo{volume}{12}~(\bibinfo{number}{4}) (\bibinfo{year}{2014})
  \bibinfo{pages}{e1001827}, \doi{\bibinfo{doi}{10.1371/journal.pbio.1001827}}.

\bibitem[{Kermack and McKendrick(1927)}]{KERMACK}
\bibinfo{author}{W.~Kermack}, \bibinfo{author}{A.~McKendrick},
  \bibinfo{title}{A contribution to the mathematical theory of epidemics},
  \bibinfo{journal}{Proc R Soc Lond B} \bibinfo{volume}{115}
  (\bibinfo{year}{1927}) \bibinfo{pages}{700--721},
  \urlprefix\url{www.plosmedicine.org}.

\bibitem[{Hethcote(2000)}]{HETHCOTE1}
\bibinfo{author}{H.~Hethcote}, \bibinfo{title}{The mathematics of infectious
  diseases}, \bibinfo{journal}{SIAM Review}
  \bibinfo{volume}{42}~(\bibinfo{number}{4}) (\bibinfo{year}{2000})
  \bibinfo{pages}{599--653}.

\bibitem[{Brauer and Castillo-Chavez(2012)}]{CCC1}
\bibinfo{author}{F.~Brauer}, \bibinfo{author}{C.~Castillo-Chavez},
  \bibinfo{title}{Mathematical Models in Population Biology and Epidemiology},
  \bibinfo{publisher}{2nd ed. Springer}, \bibinfo{year}{2012}.

\bibitem[{Bailey(1975)}]{BAILEY}
\bibinfo{author}{T.~Bailey}, \bibinfo{title}{The mathematical theory of
  infectious diseases and its applications, 2nd Ed.},
  \bibinfo{publisher}{London: Griffin}, \bibinfo{year}{1975}.

\bibitem[{Hethcote(1976)}]{HETHCOTE2}
\bibinfo{author}{H.~Hethcote}, \bibinfo{title}{Qualitative analysis of
  communicable diseases}, \bibinfo{journal}{Math Biosci}
  \bibinfo{volume}{28}~(\bibinfo{number}{4}) (\bibinfo{year}{1976})
  \bibinfo{pages}{335--356}.

\bibitem[{Jones et~al.(2008)Jones, Patel, Levy, Storeygard, Balk, Gittleman,
  and Daszak}]{JONES}
\bibinfo{author}{K.~Jones}, \bibinfo{author}{N.~Patel},
  \bibinfo{author}{M.~Levy}, \bibinfo{author}{A.~Storeygard},
  \bibinfo{author}{D.~Balk}, \bibinfo{author}{J.~Gittleman},
  \bibinfo{author}{P.~Daszak}, \bibinfo{title}{Global trends in emerging
  infectious diseases}, \bibinfo{journal}{Nature} \bibinfo{volume}{451}
  (\bibinfo{year}{2008}) \bibinfo{pages}{990--994},
  \doi{\bibinfo{doi}{10.1038/nature0653}}.

\bibitem[{Webb and Bhatia(2005)}]{WEBB}
\bibinfo{author}{P.~Webb}, \bibinfo{author}{R.~Bhatia}, \bibinfo{title}{A
  Manual: Measuring and Interpreting Malnutrition and Mortality},
  \bibinfo{publisher}{The World Food Programme WFP Nutrition Service, Rome},
  \bibinfo{year}{2005}.

\bibitem[{Solow(1956)}]{SOLOW}
\bibinfo{author}{R.~Solow}, \bibinfo{title}{Contribution to the Theory of
  Economic Growth}, \bibinfo{journal}{The Quarterly Journal of Economics}
  \bibinfo{volume}{70} (\bibinfo{year}{1956}) \bibinfo{pages}{65--94}.

\bibitem[{Cobb and Douglas(1928)}]{COBB}
\bibinfo{author}{C.~Cobb}, \bibinfo{author}{P.~Douglas}, \bibinfo{title}{A
  Theory of Production}, \bibinfo{journal}{The American Economic Review}
  \bibinfo{volume}{18}~(\bibinfo{number}{1}) (\bibinfo{year}{1928})
  \bibinfo{pages}{139--165}.

\bibitem[{Bloom et~al.(2004)Bloom, Canming, and Sevilla}]{BLOOM}
\bibinfo{author}{D.~Bloom}, \bibinfo{author}{D.~Canming},
  \bibinfo{author}{J.~Sevilla}, \bibinfo{title}{The effect of health on
  economic growth: a production function approach}, \bibinfo{journal}{World
  Development} \bibinfo{volume}{32} (\bibinfo{year}{2004})
  \bibinfo{pages}{1--13}.

\bibitem[{Moav and Neeman(2012)}]{MOAV}
\bibinfo{author}{O.~Moav}, \bibinfo{author}{Z.~Neeman}, \bibinfo{title}{Saving
  Rates and Poverty: The Role of Conspicuous Consumption and Human Capital},
  \bibinfo{journal}{The Economic Journal}
  \bibinfo{volume}{122}~(\bibinfo{number}{N563}) (\bibinfo{year}{2012})
  \bibinfo{pages}{933--956}.

\bibitem[{Levin(1998)}]{LEVIN}
\bibinfo{author}{S.~Levin}, \bibinfo{title}{Ecosystems and the biosphere as
  complex adaptive systems}, \bibinfo{journal}{Ecosystems}
  \bibinfo{volume}{1}~(\bibinfo{number}{5}) (\bibinfo{year}{1998})
  \bibinfo{pages}{431--436}, \doi{\bibinfo{doi}{10.1007/s100219900037}}.

\bibitem[{World-Bank(2015)}]{WB}
\bibinfo{author}{World-Bank}, \bibinfo{title}{Life expectancy at birth},
  \bibinfo{journal}{http://data.worldbank.org/indicator/SP.DYN.LE00.IN} .

\bibitem[{Smith et~al.(2007)Smith, McKenzie, Snow, and Hay}]{SMITH}
\bibinfo{author}{D.~Smith}, \bibinfo{author}{F.~McKenzie},
  \bibinfo{author}{R.~Snow}, \bibinfo{author}{S.~Hay},
  \bibinfo{title}{Revisiting the basic reproductive number for malaria and its
  implications for malaria control}, \bibinfo{journal}{PLoS Biol}
  \bibinfo{volume}{5} (\bibinfo{year}{2007}) \bibinfo{pages}{e42}.

\bibitem[{Becker and Hasofer(1998)}]{BECKER}
\bibinfo{author}{N.~Becker}, \bibinfo{author}{A.~Hasofer},
  \bibinfo{title}{Estimating the Transmission Rate for a Highly Infectious
  Disease}, \bibinfo{journal}{Biometrics}
  \bibinfo{volume}{54}~(\bibinfo{number}{2}) (\bibinfo{year}{1998})
  \bibinfo{pages}{730--738}.

\bibitem[{Focks et~al.(1995)Focks, Daniels, Haile, and Keesling}]{FOCKS}
\bibinfo{author}{D.~Focks}, \bibinfo{author}{E.~Daniels},
  \bibinfo{author}{D.~Haile}, \bibinfo{author}{J.~Keesling}, \bibinfo{title}{A
  simulation model of the epidemiology of urban dengue fever: literature
  analysis, model development, preliminary validation, and samples of
  simulation results}, \bibinfo{journal}{Am J Trop Med Hyg}
  \bibinfo{volume}{53}~(\bibinfo{number}{5}) (\bibinfo{year}{1995})
  \bibinfo{pages}{489--506}, \doi{\bibinfo{doi}{10.4269/ajtmh.1995.53.489}}.

\bibitem[{Medeiros et~al.(2011)Medeiros, Rodrigues, Braga, Souza, Regis, and
  Vieira}]{MEDEIROS}
\bibinfo{author}{L.~Medeiros}, \bibinfo{author}{C.~Rodrigues},
  \bibinfo{author}{C.~Braga}, \bibinfo{author}{V.~Souza},
  \bibinfo{author}{L.~Regis}, \bibinfo{author}{A.~Vieira},
  \bibinfo{title}{Modeling the Dynamic Transmission of Dengue Fever:
  Investigating Disease Persistence}, \bibinfo{journal}{PLoS Neglected Tropical
  Diseases} \bibinfo{volume}{5}~(\bibinfo{number}{1}) (\bibinfo{year}{2011})
  \bibinfo{pages}{e942}.

\bibitem[{Felipe and Adams(2005)}]{FELIPE}
\bibinfo{author}{J.~Felipe}, \bibinfo{author}{F.~Adams}, \bibinfo{title}{A
  theory of production: The estimation of Cobb-Douglas function: A
  retrospective view}, \bibinfo{journal}{Eastern Economic Journal}
  \bibinfo{volume}{31}~(\bibinfo{number}{3}) (\bibinfo{year}{2005})
  \bibinfo{pages}{427--445}.

\bibitem[{KNOEMA(2016)}]{KNOEMA}
\bibinfo{author}{KNOEMA}, \bibinfo{title}{Economy, International Comparisons,
  Labor, Productivity. Nigeria - Average depreciation rate of the capital
  stock},
  \bibinfo{journal}{https://knoema.com/PWT2015/penn-world-table-9-0?tsId=1046780}
  .

\end{thebibliography}

\end{document}